# Anomalous Charge Density Wave and Fermi Surface Reconstruction in Pressurized BaFe$_2$Al$_9$

Govindaraj Lingannan[1,2,3], M. Sundaramoorthy[2,3], Nabeel M. Jasim[4], I. K. Abbas[2], C. S. Lue[5,6], Leon F. Carstens[7], A. Bertrand[8], M. Mito[8], B. Joseph[2], Rüdiger Klingeler[7], S. Arumugam[2,9] Mahmoud Abdel-Hafiez[4,10]

[1]College of General Education, University of Doha for Science and Technology, Qatar, Doha, Qatar
[2]Centre for High Pressure Research, School of Physics, Bharathidasan University, Tiruchirappalli 620024, India
[3]Elettra-Sincrotrone Trieste S.C. p. A., S.S. 14, Km 163.5 in Area Science Park, Basovizza 34149, Italy
[4]Department of Applied Physics and Astronomy, University of Sharjah, P. O. Box 27272 Sharjah, United Arab Emirates
[5]Department of Physics, National Cheng Kung University, Tainan 70101, Taiwan
[6]Taiwan Consortium of Emergent Crystalline Materials, National Science and Technology Council, Taipei 10601, Taiwan
[7]Kirchhoff Institute of Physics, Heidelberg University, INF 227, D-69120 Heidelberg, Germany
[8]Graduate School of Engineering, Kyushu Institute of Technology, Fukuoka 804-8550, Japan
[9]Tamil Nadu Open University, Chennai 600015, India
[10]Center for Advanced Materials Research, Research Institute of Sciences and Engineering, University of Sharjah, Sharjah 27272, United Arab Emirates

## Abstract

The intermetallic compound BaFe$_2$Al$_9$ exhibits unusual physical properties associated with a charge density wave (CDW) transition. Unlike conventional CDW materials, which typically display subtle structural distortions or lattice modulations, BaFe$_2$Al$_9$ undergoes a first-order phase transition in which lattice strain plays a crucial role in the formation of the CDW state. To further explore this unique behavior, we conducted high-pressure studies, examining the electrical transport, magnetic, and structural properties to gain deeper insight into the underlying CDW mechanism. At ambient pressure, electrical resistivity and magnetization measurements confirm the presence of a CDW transition. Upon applying pressure, the CDW transition temperature (T$_{CDW}$) shifts to higher values, reaching approximately 300 K near 3.2 GPa, and the electrical resistivity increases, suggesting that pressure modulates the charge carrier concentration. Furthermore, the initially sharp first-order transition becomes more gradual, and analysis of the temperature derivative of resistivity indicates a crossover from first-order to second-order like behavior under pressure. High-pressure magnetization measurements are consistent with the electrical transport data, showing an enhancement of T$_{CDW}$ with increasing pressure. The residual resistivity increases with pressure, while the Fermi liquid coefficient A decreases above 2 GPa, pointing to a possible Fermi surface reconstruction. High-pressure synchrotron powder X-ray diffraction (XRD) measurements at room temperature reveal a lattice anomaly near 3.8 GPa, marked by a distinct trend change in macrostrain, further supporting the existence of a pressure induced structural response. These findings provide valuable insight into the nature of CDW formation in BaFe$_2$Al$_9$ and highlight the critical role of lattice strain and external pressure in tuning its electronic ground state.

# I. Introduction

Charge density wave (CDW) order is a collective electronic phenomenon commonly observed in low-dimensional solids. It arises from periodic modulations of the electronic charge density, often coupled with distortions in the atomic lattice. CDW systems have garnered extensive interest due to their complex and tuneable physical properties, such as metal-to-insulator (MI) transitions, nonlinear electrical transport, and competition or coexistence with superconductivity. A hallmark of CDW transitions is the opening of a gap at the Fermi surface (FS), which can be influenced by external parameters such as chemical substitution, strain, or pressure. Among polar intermetallics, the aluminum-rich $AM_2Al_9$ family (where A = alkaline-earth or rare-earth element, and M = transition metal) offers a fertile platform for studying CDW behavior. These compounds crystallize in a hexagonal structure with P6/mmm symmetry and host a complex electronic environment composed of intertwined kagome, honeycomb, and triangular sublattices [1,2]. Within this family, $BaFe_2Al_9$ displays a pronounced first-order phase transition below 100 K, attributed to the emergence of CDW order. In contrast, its isostructural analogs $BaCo_2Al_9$ and $SrCo_2Al_9$ remain metallic down to low temperatures and show no such anomalies. The CDW transition in $BaFe_2Al_9$ has been confirmed through electrical resistivity, magnetic susceptibility, and specific heat measurements. The resistivity shows an abrupt increase of approximately 60% at the transition temperature, while the specific heat anomaly remains relatively broad. Low-temperature powder X-ray diffraction (XRD) measurements reveal anisotropic lattice changes: a ~0.5% increase in the a-axis and a ~1.5% contraction in the c-axis, resulting in an overall unit cell volume reduction of ~0.5% near 100 K. These distortions generate an internal strain estimated at 1.5%, which may lead to mechanical instability and crystal fracture. Single-crystal XRD also confirms that the CDW in $BaFe_2Al_9$ is incommensurate [3]. The absence of similar behavior in $BaCo_2Al_9$ and $SrCo_2Al_9$ underscores the essential role of partially filled Fe 3d orbitals in driving the phase transition in $BaFe_2Al_9$. Density functional theory (DFT) calculations reveal a more complex Fermi surface and a higher density of states (DOS) at the Fermi level in $BaFe_2Al_9$ compared to its Co-based counterparts, with dominant contributions from Fe $3d$ states. The moderate electron–phonon coupling at the CDW wave vector, combined with the large DOS, suggests that electronic correlations are the primary driver of CDW formation [4,5]. Supporting evidence from $^{27}Al$ nuclear magnetic resonance (NMR) measurements includes a sudden shift in spectral features and significant reductions in both the isotropic Knight shift and the spin-lattice relaxation rate ($1/T_1$) below 100 K, indicating partial Fermi surface gapping and reduced DOS at the Fermi level [6]. Examples of three-dimensional CDW materials with first-order structural transitions include orthorhombic $U_2Co_3Si_5$-type compounds such as $Sm_2Ru_3Ge_5$ [7], $Er_2Ir_3Si_5$ [8], and $Lu_2Ir_3Si_5$ [9]. Notably, in $Er_2Ir_3Si_5$, and $Lu_2Ir_3Si_5$, the CDW transition is accompanied by a structural change from orthorhombic to triclinic symmetry. Specific heat measurements on $Er_2Ir_3Si_5$, and $Lu_2Ir_3Si_5$ show sharp peaks at the transition temperature, in contrast to the broader heat capacity enhancement observed in $BaFe_2Al_9$, although their resistivity anomalies are comparable [8,9,10]. In most CDW systems, applying pressure suppresses the transition. This suppression is typically due to the closing of the CDW gap at the Fermi surface, as pressure reduces the lattice spacing, increases orbital overlap, and alters the Fermi surface topology, thus weakening the nesting conditions essential for CDW formation [11-14]. However, a few materials show the opposite behavior [15-18], with pressure enhancing the $T_{CDW}$ and broadening the transition width. In such cases, pressure may stabilize the CDW by modifying the carrier density and electronic interactions rather than by enhancing nesting. An example is $1T-VSe_2$, where $T_{CDW}$ increases to approximately 350 K under 14.6 GPa, with a pressure coefficient of 17.8 K/GPa. High-pressure XRD measurements reveal a structural transition from trigonal to monoclinic symmetry at the same pressure, suggesting that the pressure-enhanced CDW is linked to out-of-plane Fermi surface nesting [15]. Another example is the kagome-lattice antiferromagnet FeGe, which crystallizes in the same hexagonal space group (P6/mmm) as $BaFe_2Al_9$. FeGe exhibits a pressure-enhanced CDW transition, increasing from ~110 K at ambient pressure to ~174 K at 15.2 GPa. Above 20 GPa, new diffraction peaks emerge, indicating the formation of a pressure-induced superlattice. Interestingly, in FeGe, CDW order coexists and competes with magnetism, pointing toward a non-nesting mechanism of CDW formation [17].

Motivated by these findings, we conducted a comprehensive investigation of the pressure dependence of the CDW transition in $BaFe_2Al_9$. Our study includes measurements of its electrical, magnetic, and structural properties under varying pressure conditions. We find that, unlike most CDW systems, $BaFe_2Al_9$ exhibits a pressure-enhanced $T_{CDW}$, reaching room temperature around 3.2 GPa. Analysis of low-temperature resistivity reveals Fermi-liquid behavior, and a reduction in the Fermi-liquid coefficient suggests pressure-induced Fermi surface reconstruction. High-pressure synchrotron XRD data show lattice anomalies and a clear change in microstrain near 3.8 GPa, pointing to a structural response tied to CDW tuning. These results reveal a strong coupling between electronic instability, lattice strain, and external pressure, offering new insight into the stabilization of CDW phases in three-dimensional intermetallic compounds.

## II. Experimental methods

Single crystals of BaFe$_2$Al$_9$ were grown using the aluminum self-flux method, with detailed synthesis procedures available in Ref. [4]. High-pressure electrical transport measurements were performed using a closed cycle refrigerator system, covering the temperature range of 4 K to 300 K in warming mode. To prepare samples for measurement, the grown crystals were quenched in liquid nitrogen, causing a large crystal (~2 mm) to fracture into smaller pieces (ranging from ~50 to 400 μm). From these, a well-shaped rectangular sample was selected for electrical transport measurements. Initially, we attempted measurements using a relatively large sample (~1 mm), but upon cooling below 100 K, the sample fractured into two pieces, leading to an open-circuit error in our system. Subsequently, we gradually reduced the sample size, and optimal results were obtained with a sample 350 μm in length. High pressure electrical transport measurements were conducted using a clamp-type hybrid double-cylinder piston pressure cell, with pressure cell preparation details provided in Ref. [19]. Ambient and high pressure Powder XRD measurements were carried out at the XPRESS beamline of Elettra Sincrotrone Trieste, using a Pilatus 6M large-area detector. To account for instrumental contributions to peak broadening, the beamline setup was calibrated using a standard CeO$_2$ (SRM674b- National Institute of Standard Technology) reference sample. The instrumental profile parameters (U, V, W) and sample displacement/zero correction parameters (X, Y, Z) were extracted and subsequently fixed in GSAS-II. These calibrated values were then applied during Rietveld refinement to ensure accurate modelling of the diffraction profiles. Detailed experimental procedures are described in Ref. [19]. Magnetic measurements were performed using a Quantum Design MPMS3 SQUID magnetometer. For high pressure magnetization studies, a nonmagnetic BeCu piston-cylinder pressure cell was employed. All magnetization measurements were conducted under an applied magnetic field of 1 Tesla [20]. Static magnetic susceptibility measurements were performed under an applied magnetic field of 0.1 T.

## III. Results and discussion
### a. High pressure electrical resistivity

Figure 1 presents the temperature-dependent electrical resistivity of the BaFe$_2$Al$_9$ single crystal under various applied pressures. At ambient pressure, a distinct resistivity drop is observed near 112 K, marking the CDW transition temperature, as also shown in Fig. S1(a). The value of $T_{CDW}$ was precisely determined by taking the first derivative of resistivity with respect to temperature (d$\rho$/d$T$), shown in Fig. S2(b). The relative resistivity change at the transition, calculated using the relation $\Delta\rho/\rho(T_{CDW})$, yields approximately 67%, consistent with previously reported values [Refs. 3, 4]. At ambient conditions, the resistivity displays metallic behavior down to ~110 K, followed by a sharp drop due to the onset of CDW order. Above $T_{CDW}$, the resistivity exhibits a linear temperature dependence, attributed primarily to electron–phonon scattering, as indicated by green guideline fits in Fig. 1. Upon applying pressure, $T_{CDW}$ systematically shifts to higher temperatures, with an estimated positive pressure coefficient of d$T_{CDW}$/d$P \approx$ 60 K/GPa, the highest reported pressure response for a CDW transition to date. $T_{CDW}$ values extracted from the d$\rho$/d$T$ analysis at each pressure are summarized in Fig. S2. As pressure increases, the CDW transition becomes more gradual, reflecting a crossover from a sharp, first-order to a broadened, second-order-like transition [21]. This broadening is clearly visible in the derivative plots and resistivity curves, indicating a possible change in the underlying transition mechanism. Figure 1(b) shows that $T_{CDW}$ increases linearly up to ~3 GPa, and that the transition width becomes significantly broader with pressure. Additionally, the data suggest a subtle change in lattice compression near 3.2 GPa at 300 K, possibly hinting at pressure-induced structural or electronic modifications.

To further explore the low-temperature electronic behavior, we fitted the resistivity curves using the Fermi liquid (FL) model: $\rho(T)=\rho_0+AT^2$, where $\rho_0$ is the residual resistivity and $A$ is the FL coefficient, which reflects electron–electron scattering strength. The fits, shown as red lines in Fig. 1, apply well up to 100 K at 0 and 0.5 GPa. At higher pressures, the fitting range extends in correspondence with the pressure-induced increase of $T_{CDW}$, fitting up to ~110 K at 1.0–2.0 GPa, ~120 K at 2.5 GPa, and ~130 K at 3.0 GPa. The extracted fit parameters are listed in Table 1. The Kadowaki–Woods (KW) ratio, given by $A/\gamma_e^2$, was calculated to assess correlation strength. Here, A is the FL coefficient obtained from the fit, and $\gamma_e$ is the Sommerfeld coefficient of the electronic specific heat ($\gamma_e$ = 19.4 mJ mol$^{-1}$ K$^{-2}$, from Ref. [3]). Our $A/\gamma_e^2$ values increase with pressure and remain close to the universal KW value, indicating that BaFe$_2$Al$_9$ behaves as a strongly correlated Fermi liquid within the studied pressure range [22]. The pressure dependence of $\rho_0$ shows two distinct trends: a moderate increase below 1 GPa, followed by a more pronounced rise above 1.5 GPa (Fig. S3). The FL coefficient A increases with pressure up to ~2 GPa, suggesting strengthening electron–electron correlations, but decreases at higher pressures. This non-monotonic behavior may be attributed to pressure-induced microstrain or structural disorder, which could reduce correlation effects by disturbing the electronic coherence. These results highlight a complex interplay between pressure, microstructural strain, CDW formation, and scattering mechanisms in BaFe$_2$Al$_9$. The enhanced residual resistivity and evolving FL parameters collectively point toward Fermi surface reconstruction and the critical role of lattice effects in shaping the CDW ground state under pressure.

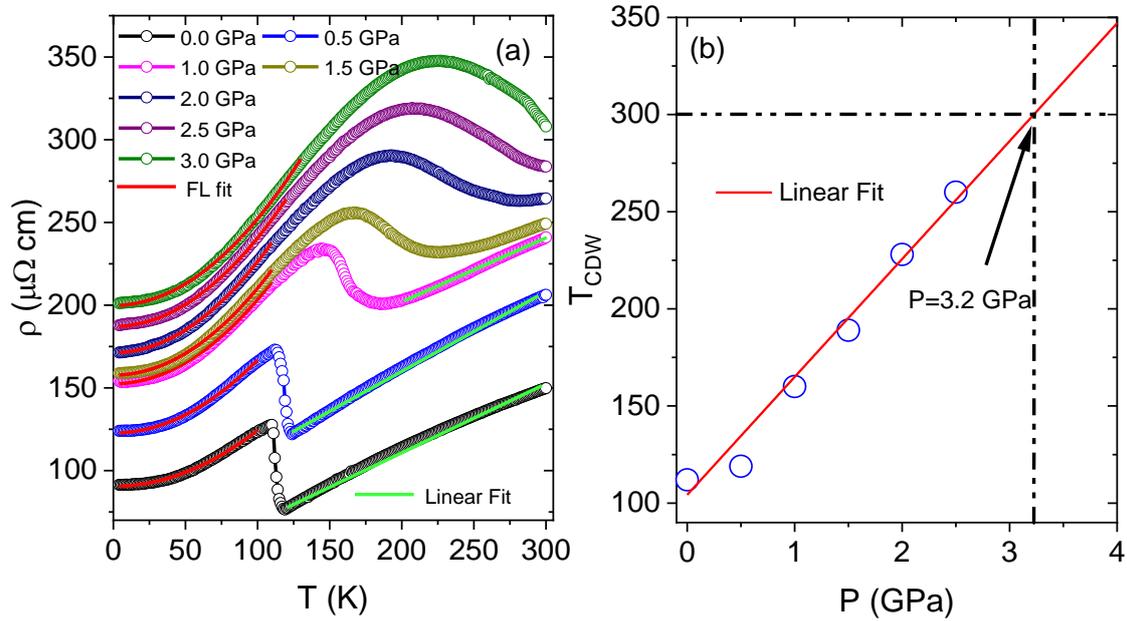

*Figure 1: (a) Temperature dependence of electrical resistivity measurements under various hydrostatic pressures, (b) $T_{CDW}$ as a function of Pressure.*

| P (GPa) | $\rho_0$ (μΩ cm) | A (μΩ cm) | $A/\gamma_e^2 \times 10^{-5}$ (μW cm mol$^2$ K$^2$ mJ$^{-2}$) |
|---|---|---|---|
| 0 | 90.47429 | 0.0034 | 0.90 |
| 0.5 | 122.73981 | 0.00441 | 1.17 |
| 1 | 152.46971 | 0.00509 | 1.35 |
| 1.5 | 157.71752 | 0.00525 | 1.39 |
| 2 | 171.28392 | 0.00551 | 1.46 |
| 2.5 | 187.11208 | 0.00535 | 1.42 |
| 3 | 199.80144 | 0.00524 | 1.39 |

*Table 1: Pressure-dependent PL fitting parameters for BaFe$_2$Al$_9$. The table lists the applied pressure (P), residual resistivity ($\rho_0$), FL coefficient (A), and the calculated KW ratio. These parameters reflect the evolution of electron–electron scattering and correlation strength under pressure.*

### b. High pressure magnetization

The static magnetic susceptibility shown in Fig. S4 (a) does not significantly depend on temperature except for a sharp jump centered around $T_{CDW}$ = 107(1) K and a Curie-like upturn at low temperatures. The only weak temperature dependence suggests a Pauli-paramagnetic behavior with negligible changes of the DOS consistent with a recent report [3]. The strong jump is indicative of a discontinuous phase transition and the associated changes in the static susceptibility of $\Delta\chi \simeq 3.4 \times 10^{-4}$ erg/(G$^2$mol) indicate considerable changes of the Fermi surface at $T_{CDW}$. At low-temperatures, the Curie-like tail signals the existence of quasi-free localized moments. The presence of such moments is proven by the isothermal magnetization, at 1.8 K, which features an overall linear behavior superimposed by a Brillouin-like response (Fig. S4 (b)). Both fitting the data and extrapolating the linear response at high fields to zero field implies a saturation moment of $M_{sat}^{qf} = 1 \times 10^{-2}$ μ$_B$/f.u. associated with quasi-free moments, i.e., moments obeying a Brillouin-like field dependence which can be aligned in rather small magnetic fields.

Figure 2(a) shows the temperature dependent magnetization of BaFe$_2$Al$_9$ measured under various fixed pressures up to 1.4 GPa. The data were collected during the heating process under an applied magnetic field of 1 T. Figure 2(b) presents the evolution of the onset of $T_{CDW}$ as a function of pressure. With increasing pressure, $T_{CDW}$ shifts linearly to higher temperatures, consistent with the trend observed in pressure-dependent electrical transport measurements. Figure S5 compares the magnetization data during cooling and heating at ambient pressure, clearly confirming the presence of a CDW transition with thermal hysteresis (~10 K). Specifically, at ambient pressure, $T_{CDW}$ occurs near 98 K during cooling and shifts to approximately 107 K during heating, consistent with a first-order phase transition. From 300 K down to ~160 K, the magnetization remains nearly flat, indicating the absence of long-range magnetic ordering. Below 150 K, a slight upturn in magnetization is observed, likely originating from paramagnetic impurities or localized Fe moments that become weakly active at low temperatures. The systematic pressure-dependent magnetization data shown in Fig. 2(a), collected during the heating process, reveal that $T_{CDW}$ increases with pressure, reaching approximately 273 K at 1.4 GPa. After pressure is released, the Magnetization data were recovered to the initial one, suggesting that nice hydrostatic condition was maintained below 1.4 GPa. Complementary data collected during the cooling process are shown in Fig. S6. The M(T) signal is relatively weak, which makes it challenging to accurately determine the onset, completion, and thermal hysteresis of the $T_{CDW}$ transition. At higher pressures, only the onset could be identified with confidence. The slight differences between the $T_{CDW}$ values obtained from magnetization and electrical resistivity measurements result from the use of different criteria: in electrical transport, $T_{CDW}$ is defined by the peak in the first derivative of resistivity (dρ/dT), while in magnetization it is identified by the initial appearance of the anomaly.

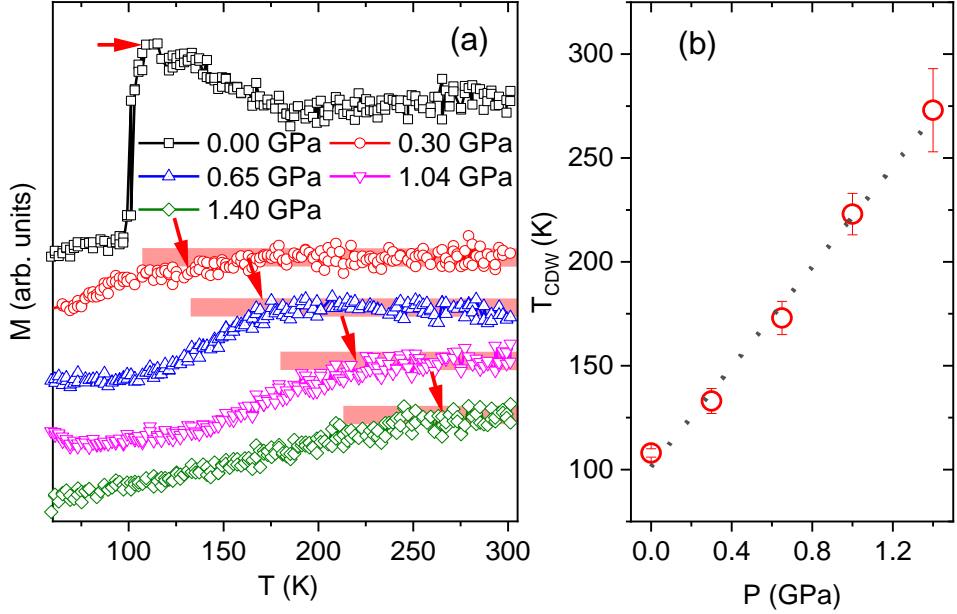

*Figure 2: (a) Temperature dependent magnetization of BaFe$_2$Al$_9$ under various fixed pressures, measured during the heating process. (b) T$_{CDW}$ as a function of Pressure.*

### c. High pressure synchrotron powder XRD

We initially anticipated a first-order phase transition associated with the T$_{CDW}$ near 3.2 GPa at 300 K, as suggested by high-pressure electrical transport measurements. To investigate this, we performed pressure-dependent synchrotron powder X-ray diffraction (PDXRD) experiments at room temperature. Figure S7 presents the collected PDXRD patterns of BaFe$_2$Al$_9$ up to 7.02 GPa. No appearance or disappearance of diffraction peaks was observed across this pressure range, indicating the absence of a structural phase transition. Only a slight shift of the diffraction peaks toward higher angles with increasing pressure was detected, consistent with lattice contraction. At ambient pressure, several diffraction peaks, specifically (200)/(101), (210)/(201), (220)/(301), and the overlapping (002)/(411)/(222) are closely spaced. As pressure increases, these peaks gradually move apart, which is indicative of anisotropic lattice compression. The structural symmetry remains preserved throughout the pressure range, and these peak separations are highlighted by blue arrows in Fig. S7. Figure S8 (a–d) shows Rietveld refinements of the diffraction data at selected pressures: 0 GPa, 2.52 GPa, 5.10 GPa, and 7.02 GPa. The refinements were performed using the CIF file from Ref. [3]. No additional diffraction peaks nor the disappearance of existing peaks were observed, supporting the conclusion that no structural phase transition occurs up to 7.02 GPa. Figure S9 displays the pressure dependence of the lattice parameters (a and c) and unit cell volume. No evidence of a first-order transition in lattice parameter is observed near the expected pressure of 3.2 GPa.

Instead, we observe a pressure-induced change in the compression trends of the lattice parameters. In the low-pressure regime, the lattice parameter a compresses at a rate of −0.017 Å/GPa, whereas at higher pressures, this rate reduces to −0.011 Å/GPa. For the c-axis, the compression rate is −0.029 Å/GPa at low pressure and −0.028 Å/GPa at high pressure, showing a much smaller variation compared to a. This confirms that the structure is more compressible in-plane than out-of-plane, demonstrating pressure-induced anisotropy. With increasing pressure, the faster compression of a relative to c leads to a decreasing c/a ratio, as shown in Fig. 3(a). However, the decrease is nonlinear, indicating a complex anisotropic response of the crystal lattice. Since the Fe–Al sublattice resides mainly in the basal plane, the pronounced in-plane compression suggests potential pressure-induced distortion or instability within the Fe–Al framework. Despite these internal lattice changes, no new or missing diffraction peaks are detected, again confirming the absence of a crystallographic phase transition. Further evidence for pressure-induced anisotropy is supported by both lattice parameter trends and bulk modulus analysis. The bulk modulus ($B_0$) and zero-pressure volume ($V_0$) were extracted by fitting the unit cell volume to a third-order Birch–Murnaghan equation of state, yielding values of $B_0$ = 75.0 GPa and $V_0$ = 219.8 Å$^3$. Additionally, microstrain (μ-strain) was refined using the GSAS-II software. To quantify anisotropic lattice distortions in BaFe$_2$Al$_9$ under pressure, we employed Stephens' phenomenological model for anisotropic peak broadening, implemented within the Rietveld refinement framework [23]. Due to the hexagonal symmetry of the compound (space group P6/mmm), the strain-induced broadening of Bragg reflections was modeled as a function of the Miller indices *hkl*, using the variance of the reciprocal lattice spacing $M_{hkl}=1/d_{hkl}^2$. For hexagonal systems, this variance is given by:

$$\text{Var}(M_{hkl}) = S_{400}(h^2+hk+k^2)^2 + S_{202}(h^2+hk+k^2)l^2 + S_{004}(l)^4.$$

Here, $S_{400}$, $S_{202}$, and $S_{004}$ are the anisotropic strain coefficients corresponding to in-plane, mixed, and out-of-plane strain components, respectively. These parameters were extracted from HP-XRD data.

Figure 3: (b–d) shows the calculated microstrain model parameters $S_{004}$, $S_{202}$, and $S_{400}$ plotted as a function of pressure using GSAS-II [24]. Our analysis reveals that both $S_{400}$ and $S_{004}$ increase systematically with pressure, indicating enhanced microstrain along both the a–b plane and the c-axis. In contrast, the mixed component $S_{202}$ decreases with pressure. Importantly, all three coefficients exhibit a clear trend change near 3.8 GPa, closely aligning with the anomaly observed in electrical transport measurements at 3.2 GPa, where a broadened CDW transition

was detected near 300 K. This correspondence strongly suggests a coupling between anisotropic lattice strain and the evolution of the CDW phase under pressure. The observed increase in directional microstrain likely facilitates electronic instability and may play a critical role in stabilizing the high-pressure CDW state in $BaFe_2Al_9$.

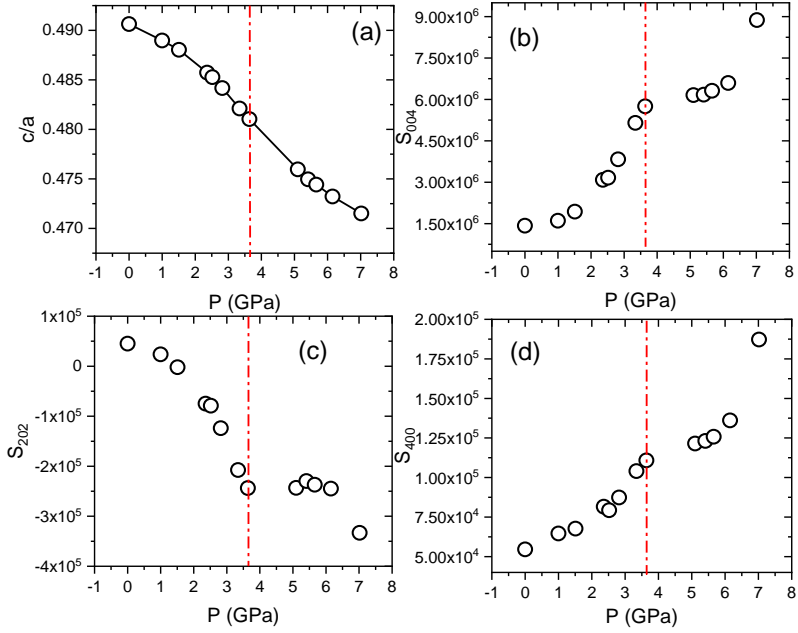

Figure 3: (a) c/a ratio, and (b–d) microstrain model parameters $S_{004}$, $S_{202}$, and $S_{400}$ plotted as a function of pressure.

## IV. Conclusion

We have systematically investigated the electrical transport, magnetic, and structural properties of $BaFe_2Al_9$ under applied pressure to elucidate the nature and evolution of its CDW transition. At ambient pressure, electrical resistivity and magnetization measurements confirm the presence of a first-order CDW transition. With increasing pressure, the resistivity is enhanced, indicating a possible reconstruction of the Fermi surface. Simultaneously, the first-order transition becomes broader and shifts steadily to higher temperatures, with the CDW transition temperature reaching approximately 300 K near 3.2 GPa. The residual resistivity, extracted from Fermi-liquid fitting, exhibits two distinct pressure-dependent trends, suggesting a crossover in the electronic structure. Furthermore, the FL coefficient decreases above 2 GPa, implying a reduction in electron–electron correlations and supporting the scenario of a pressure-induced Fermi surface reorganization. High pressure magnetization measurements reinforce these findings, showing a consistent enhancement of $T_{CDW}$. Synchrotron HP-XRD measurements up to 7.02 GPa at room temperature do not reveal a structural phase transition but do show a subtle anomaly in lattice parameters near 3.8 GPa. The lattice exhibits anisotropic compression, which likely contributes to the evolution of the CDW state. Importantly, a pronounced anomaly in the calculated anisotropic microstrain coincides with the pressure region where $T_{CDW}$ approaches 300 K, further linking microstrain to CDW stabilization. The absence of any discontinuous change in lattice parameters with pressure, combined with the gradual increase in microstrain and the broadening of the transport anomaly, suggests that the CDW transition evolves from first-order at ambient pressure to a more second-order-like character under high pressure. This behavior is consistent with a continuous structural response in which the order parameter develops smoothly, and no latent heat is involved. The role of microstrain in driving this transition underscores the strong coupling between lattice fluctuations and the electronic instability responsible for CDW formation. These results provide a comprehensive understanding of the pressure induced tuning of the CDW transition in $BaFe_2Al_9$. The transition evolves from first-order to second-order-like character under pressure, with microstrain playing a central role in this transformation. In addition, the observed Fermi surface reconstruction above 2 GPa highlights a strong coupling between lattice strain and the electronic degrees of freedom in this system.

**ACKNOWLEDGMENT**

N.J. and MAH acknowledge the Advanced Materials Research Lab at the University of Sharjah. The authors thank the Xpress beamline of the Elettra Sincrotrone Trieste for beamtimes (Proposal No: 20220489). G.L., M.H and acknowledges the support from ARG01-0516-230179 project funded by QRDI. G.L., and M.S. gratefully acknowledge the receipt of a fellowship from the ICTP Programme for Training and Research in Italian Laboratories (TRIL), Trieste, Italy. S.A. wishes to thank DAE-BRNS (Mumbai) for their financial support. R.K. and L.F.C. acknowledge support by the DFG via the Heidelberg STRUCTURES Cluster (EXC2181/1-390900948).

# Supporting information

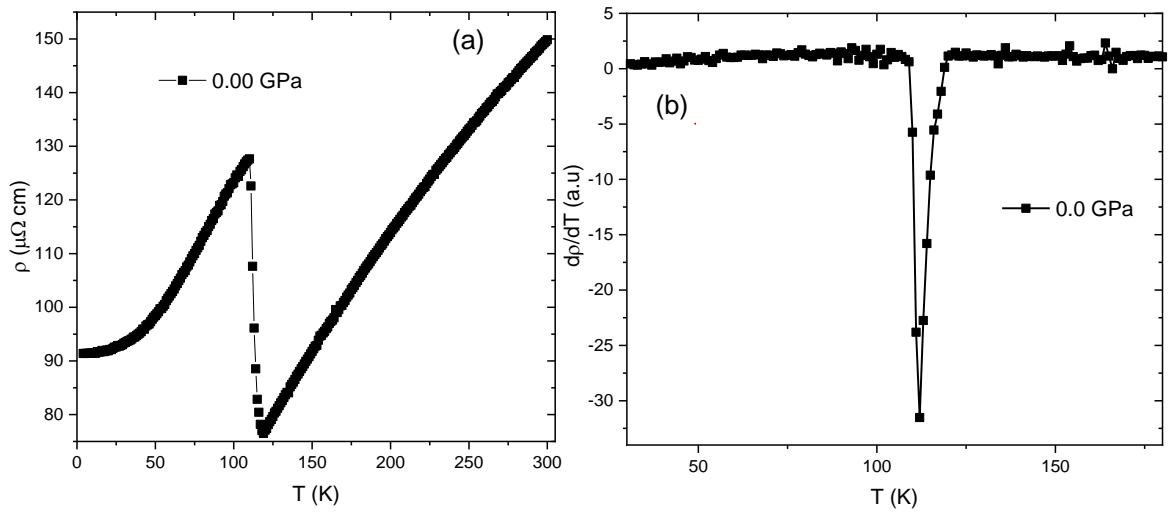

*Figure S4:(a) Temperature dependence of electrical resistivity under 0 GPa pressure, (b) b) First derivative of the temperature dependence of electrical resistivity demonstrating the CDW transition temperature*

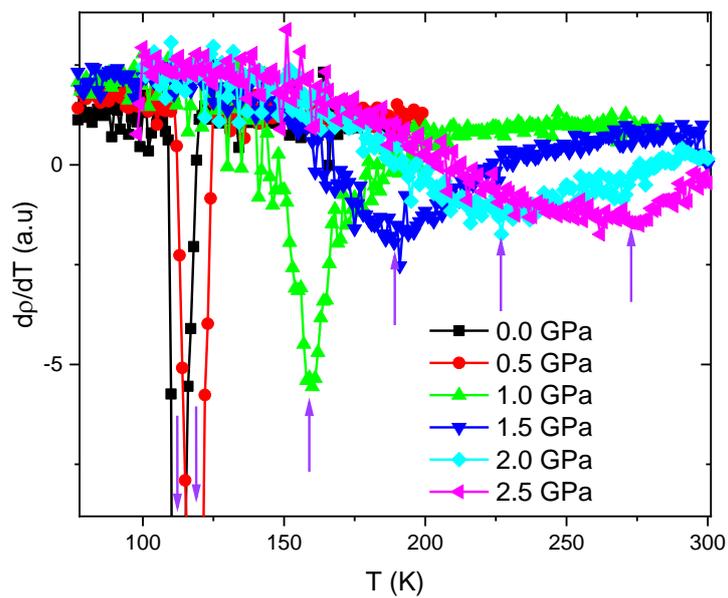

*Figure S5. First derivative of the temperature dependence of electrical resistivity under various pressure.*

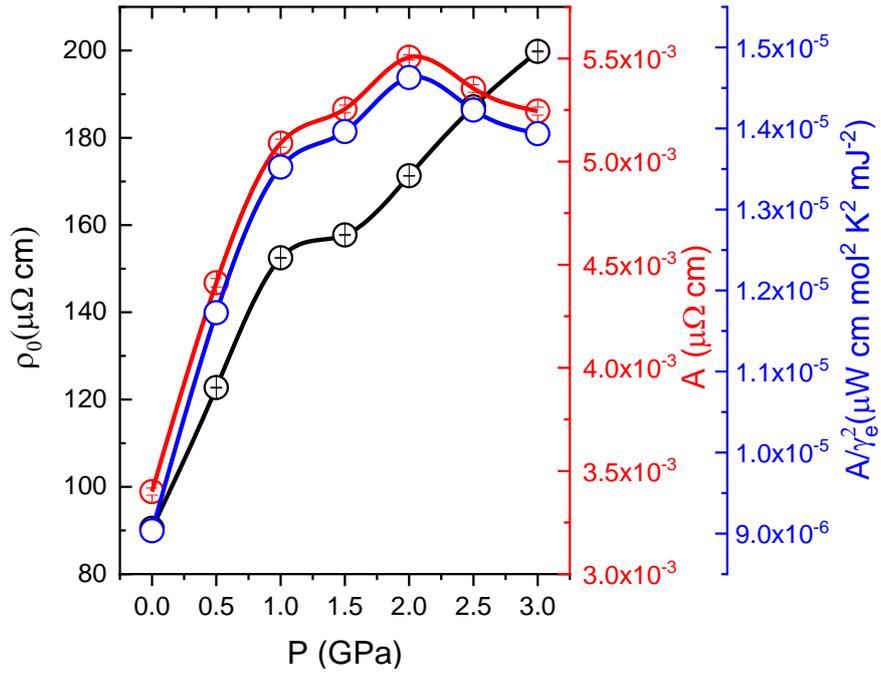

Figure S6: Residual resistivity, Fermi liquid coefficient and KW ratio as a function of pressure

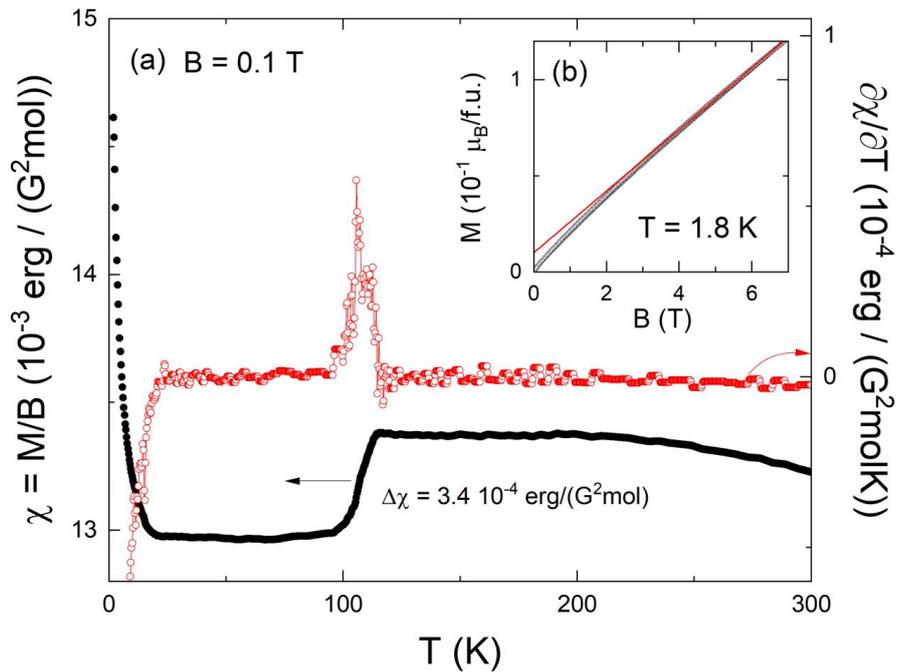

Figure S7: Static magnetic susceptibility $\chi = M/B$, at $B = 0.1$ T (left ordinate), and its derivative $\partial \chi / \partial T$ (right ordinate). Inset: Isothermal magnetization at $T = 1.8$ K. The red line extrapolates the behavior at 7 T linearly to zero field suggesting the presence quasi-free moments saturating at $M_{sat}^{qf} = 1 \times 10^{-2}$ μB/f.u.

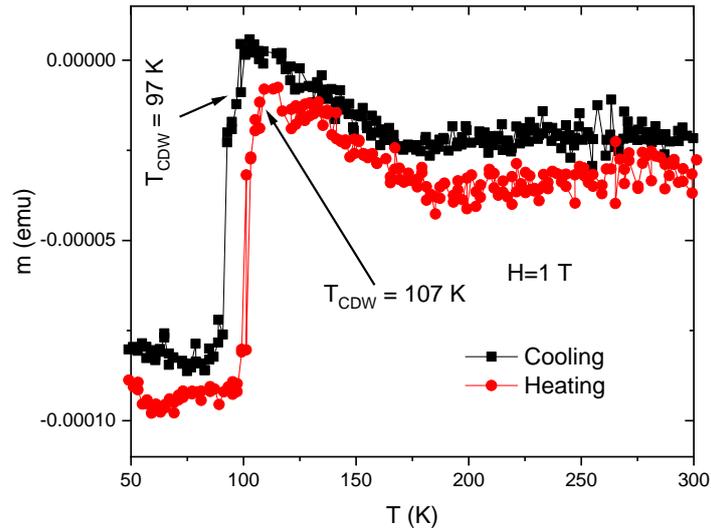

*Figure S8: Temperature-dependent magnetization at P = 0 GPa under a magnetic field of 1 T. Data collected during the cooling process are shown in black, while data collected during the warming process are shown in red.*

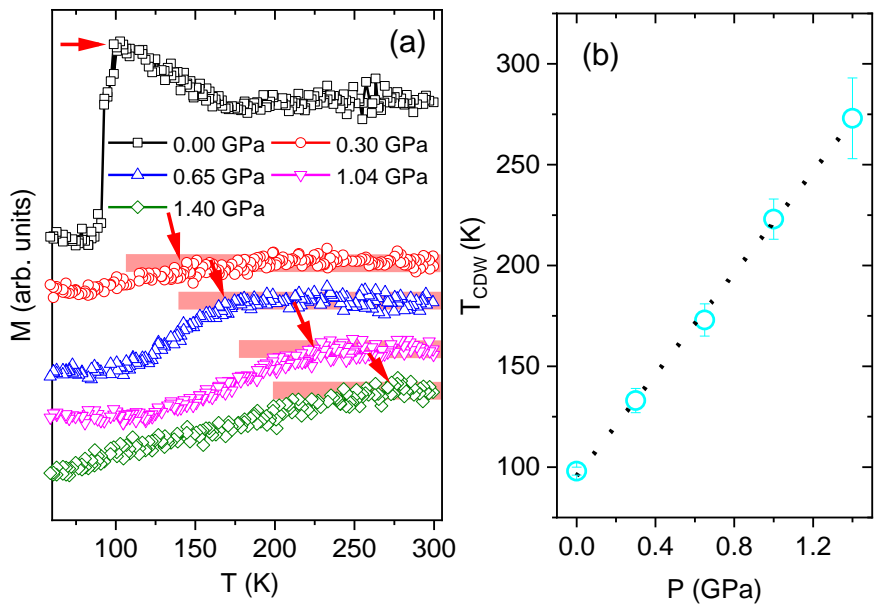

*Figure S9: (a) Temperature dependent magnetization of BaFe$_2$Al$_9$ measured under various fixed pressures during the cooling process. (b) T$_{CDW}$ as a function of Pressure, showing both cooling and heating process data for comparison.*

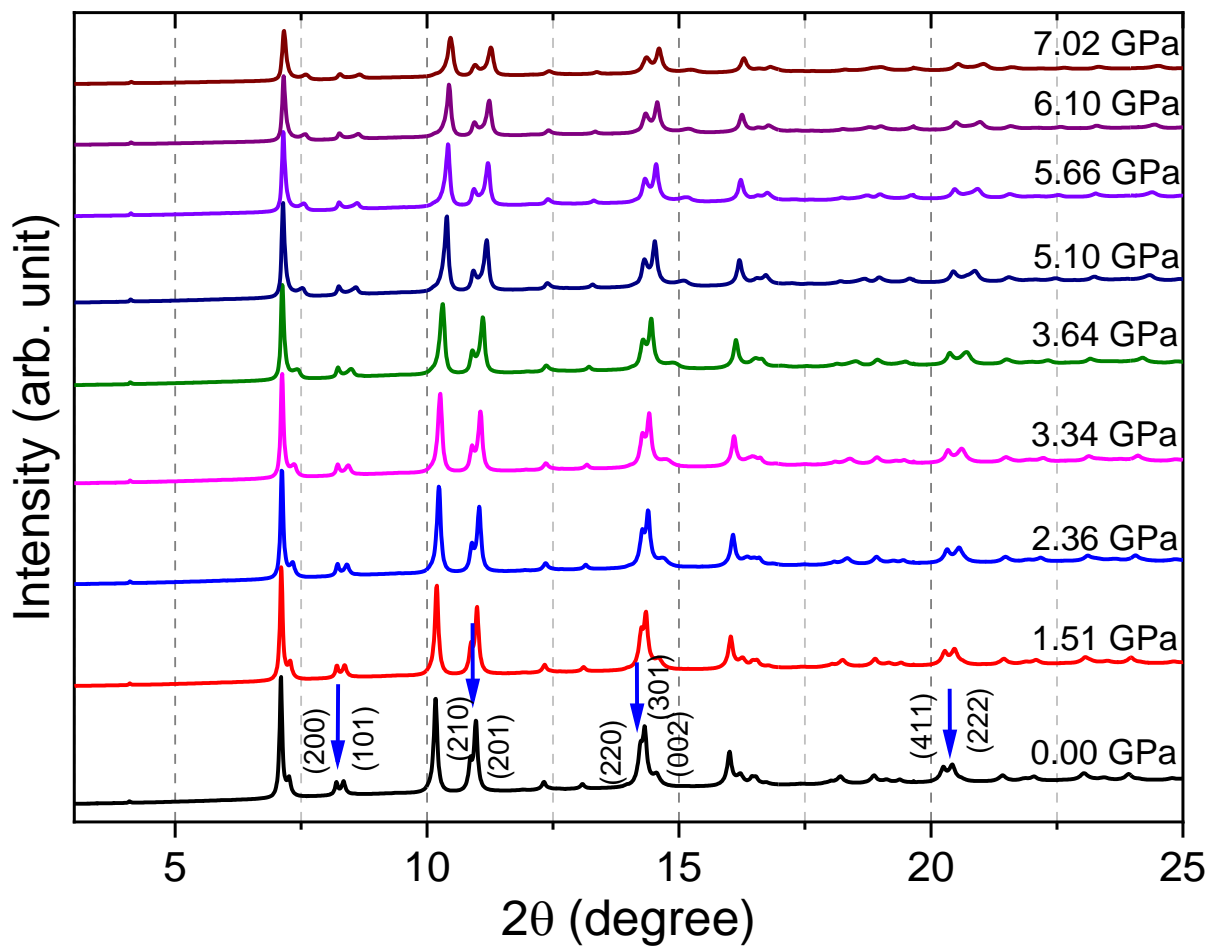

Figure S10: Measured x-ray-diffraction pattern of BaFe$_2$Al$_9$ at several pressures up to 7.02 GPa using 4:1 methanol-ethanol as pressure transmitting medium. Patterns are shifted vertically for clarity in presentation.

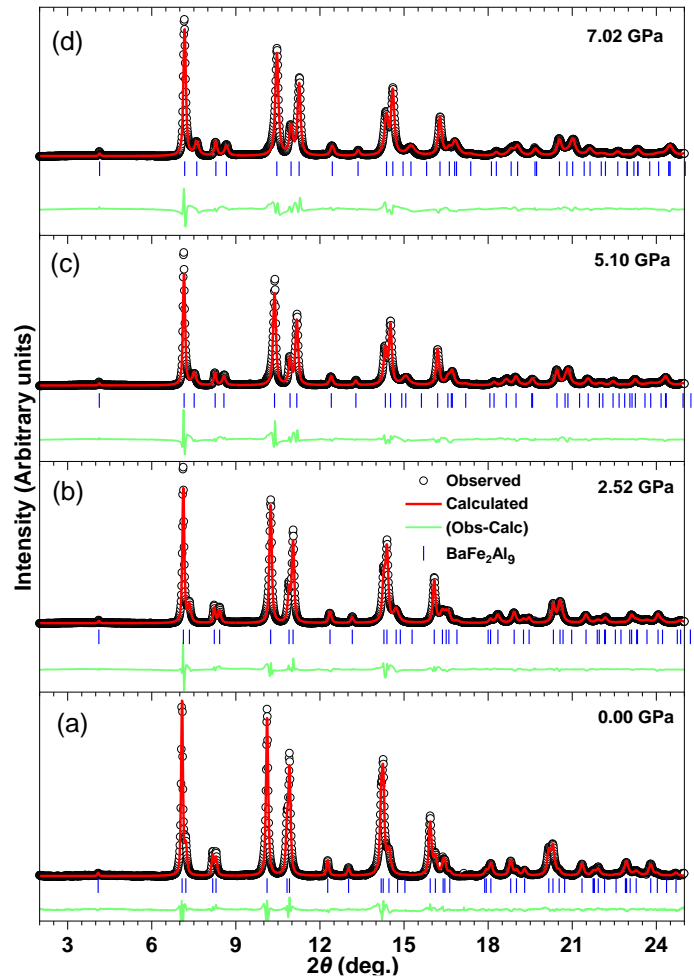

*Figure S11:* Measured x-ray-diffraction pattern of BaFe$_2$Al$_9$ sample (a) 0.00 GPa, (b) 2.52 GPa, (c) 5.10 GPa and (d) 7.02 GPa together with the Rietveld refinement results (solid red lines). Blue vertical bars below the diffractogram indicate the Bragg peak positions.

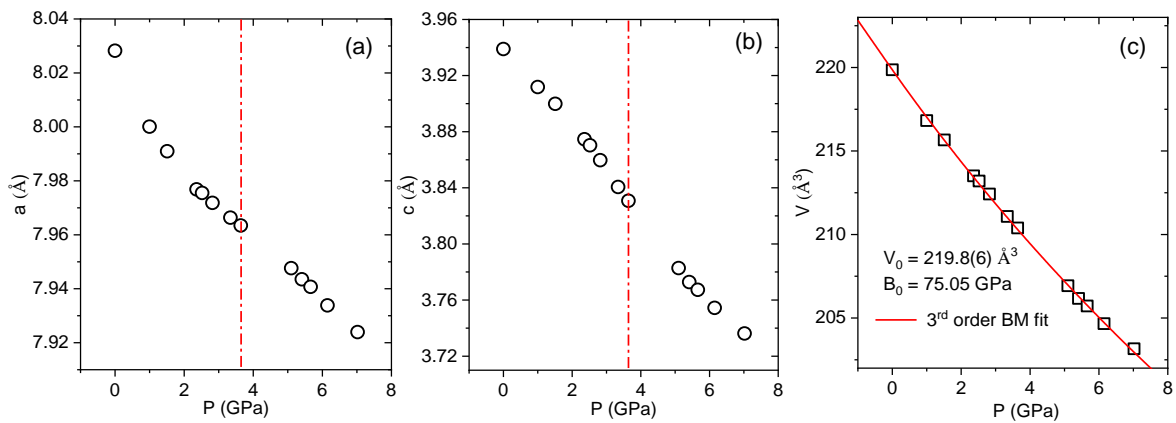

Figure S12: *Pressure dependence of lattice parameters a (a), c (b), and unit cell volume (c).*